# Description of Four- and Five-Nucleon Systems by Solving Faddeev-Yakubovsky Equations in Configuration Space

*Rimantas Lazauskas[1]\* and Jaume Carbonell[2]*

[1] IPHC, IN2P3-CNRS/Université de Strasbourg BP 28, Strasbourg, France, [2] Université Paris-Saclay, CNRS/IN2P3, IJCLab, Orsay, France

The Faddeev-Yakubovsky equations constitute a rigorous formulation of the quantum mechanical N-body problem in the framework of non-relativistic dynamics. They allow the exact solutions of the Schrödinger equation for bound and scattering states to be obtained. In this review, we will present the general formalism as well as the numerical tools we use to solve Faddeev-Yakubovsky equations in configuration space. We will consider in detail the description of the four- and five-nucleon systems based on modern realistic nuclear Hamiltonians. Recent achievements in this domain will be summarized. Some of the still controversial issues related with the nuclear Hamiltonians as well as the numerical methods traditionally employed to solve few-nucleon problems will be highlighted.

**Keywords: Faddeev-Yakubovsky equations, four-nucleon system, five-nucleon systems, few-body collisions, scattering observables**

## 1. INTRODUCTION

The solution of the Faddeev-Yakubovsky (FY) equations is an extremely challenging task from both the intellectual and technical points of view. The fast growth in the complexity of this problem with the number of interacting nucleons (A) makes progress in solving these equations relatively slow [1]. During the last twenty years, we have witnessed the emergence of the full solution—bound and scattering states—of the four-nucleon problem [2, 3] and only very recently have the first solutions for A = 5 [4–6] been published.

Although the four-boson bound problem was already formulated—and solved with S-wave interaction—in [2], the first converged result employing realistic NN interactions for the A = 4 bound state (⁴He) took another 10 years to achieve [3].

The first solution of the scattering problem for the elastic 1+3, 2+2, and 1+3→ 2+3 rearrangement channels within the isospin approximation and S-wave interactions dated from 1998 [7], and it took twenty years more to obtain a full solution of the four-nucleon scattering problem with (i) realistic interactions [8–10] (ii) including charge-dependent (CD) and non local terms [11], (iii) Coulomb effects [12–15], (iv) three- and four-body breakup amplitudes [13, 16–18], and (v) a proper *ab-initio* determination of the 4N resonant states (e.g., 4n or ⁴H) as S-matrix poles in the complex energy plane [6, 19–23]. There remains only the computation and analysis of the three- and four-body breakup differential cross-sections, since only the integrated cross-sections are nowadays available.

The A = 4 schematic chart is displayed in the left panel of **Figure 1**. It comprises five different charge states (Z = 0, 1, 2, 3, 4), including a single bound state ($^4$He) as well as five two-cluster scattering channels (n-$^3$H, n-$^3$He, p-$^3$H, $^2$H-$^2$H, $^2$H-$^3$He denoted in an olive color), several three-body (in blue) and four-body (in black) break-up thresholds, and numerous—well-identified or questioned—low-energy resonances. The A = 4 sector presents the simplest case, revealing in practice all the phenomena of the theoretical nuclear physics: the presence of several thresholds and resonances. As an example, the continuum of the $^4$He nucleus contains almost degenerate n-$^3$He and p-$^3$H thresholds, with an $^4$He resonant state situated in between whose position must be accurately determined since it strongly modifies the scattering in both channels and thus constitutes a serious challenge for all realistic NN interaction models [8, 26]. Although still far from the intricacy of heavy nuclei, one can say that, in some sense, the nuclear complexity really starts at A = 4.

Solving the A = 5 problem represents a redoubtable technical and numerical difficulty with respect to the A = 4 case. However, the A = 5 chart is simpler than the A = 4 one due to the absence of the A = 5 bound state and of $^{3,5}$Li and/or $^4$Be targets: the number of charge states effectively investigated is limited to three (Z = 1, 2, 3) since the experimentally inaccessible 5n and 4p states raise less interest. There are four two-body scattering channels (n-$^4$He, $^2$H-$^3$H, $^2$H-$^3$He, and p-$^4$He, also denoted in olive), some of them, like the $^2$H+$^3$H → n+$^4$He fusion reaction, of paramount importance in nuclear physics and in the stellar nucleosynthesis cycle. In contrast, the number of three- and four-body breakup thresholds (in blue) is sensibly larger. This is illustrated in the right panel of **Figure 1**, where the "nuclear chart" corresponding to A = 5 is displayed. The FY solutions for A = 5 are at present limited to low energy (S- and P-waves) n-$^4$He elastic scattering [4, 5] and in computing the lowest resonant states of $^5$H [6], in both cases using realistic interactions. Some disagreements with the R-matrix analysis were found in both systems. The fusion reaction $^2$H+$^3$H → n+$^4$He has not yet been solved within the FY framework, but a recent pioneering result has been achieved within the NCSMC approach [27].

We would like to point out from the very beginning that other rigorous schemes were proposed for solving the *ab initio* N-body problem. One of the most relevant is that provided by the AGS equations [28], which is strictly equivalent to the FY formalism and has produced very accurate results for the three- and four-nucleon problem, always in momentum space [29, 30]. We also emphasize that such rigorous mathematical schemes are not necessary when dealing with bound states or simple 1 + (N − 1) elastic scattering processes and that the Schrödinger equation can then be directly solved by several methods.

It is worth also mentioning that, aside from FY solutions in configuration space, on which we report, there are several competing approaches to solving the A = 4 and A = 5 problems that have produced very interesting and, in some cases, pioneering results. Any attempt at an exhaustive reference list is beyond our capabilities. However, we would like to point out among them the GFMC [31, 32], variational approaches with Hyperspherical Harmonics [33, 34] or a Gaussian basis [35], RGM [36], NCSM and NCSMC [27, 37], and Lorentz

Integral Transform [38], which can produce very accurate results, in some cases well beyond the technical capabilities of the Faddev-Yakubovsky approach. However, the Faddeev-Yakubovsky partition of the wave function is interesting for increasing the numerical convergence of the results or is even unavoidable for an appropriate implementation of the boundary conditions [35, 39]. The interested reader can find a more thorough bibliography in some devoted reviews [40].

In this contribution, we will concentrate on some particular issues that our previous works had not treated with the required detail. We will mostly present results related to the four-nucleon scattering problem, obtained by solving the Faddeev-Yakubovsky equations in configuration space, and will add some recent results on the five-nucleon n-$^4$He low-energy scattering. In section 2, we will detail the theoretical aspects of the four-body equations. section 3 is devoted to the discussion of 4N scattering results with different realistic models. Some concluding remarks are collected in the Conclusions.

## 2. THEORETICAL DESCRIPTION

In what follows, we will present the general formalism as well as the numerical methods relevant to the solution of the four-body problem in configuration space. Some results related with neutron scattering on $^4$He that we obtained by solving five-body FY equations will also be discussed; however, and due to its complexity, the five-body formalism will not be presented here. For this particular case, a interested reader may refer to Sasakawa [41], Lazauskas [4, 5], and Lazauskas and Song [42].

### 2.1. The Four-Body FY Equations
The derivation of the four-body Faddeev-Yakubovsky equations starts by defining the three-body-like Faddeev components (FC) $\psi_{ij}$, which are associated to each interacting pair of particles (ij):

$$\psi_{ij} = G_0 V_{ij} \Psi \quad (i < j).$$ (1)

Here, $G_0 = (E − H_0)^{-1}$ denotes the free four-body Green's function, associated with the four-body kinetic energy operator $H_0$ and the four-body energy $E$, while $V_{ij}$ denotes the two-body potential between the particles $i$ and $j$. Naturally, for a four-body system, there exist six different three-body-like Faddeev components. In terms of these, one can define two types of the so-called Faddeev-Yakubovsky component (FYCs), denoted, respectively type-K and type-H components, by the relations:

$$\begin{aligned} \mathcal{K}^l_{ij,k} &= G_{ij} V_{ij} (\psi_{jk} + \psi_{ik}) \quad (i < j); \\ \mathcal{H}^{kl}_{ij} &= G_{ij} V_{ij} \psi_{kl} \quad (i < j; \quad k < l). \end{aligned}$$ (2)

In this equation, $G_{ij} = (E − H_0 − V_{ij})^{-1}$ denotes the interacting four-body Green's function associated with the interaction term between particles $i$ and $j$. By permuting particle indexes, one may construct 12 independent components of type-K as well as six independent components of type-H. The asymptotes

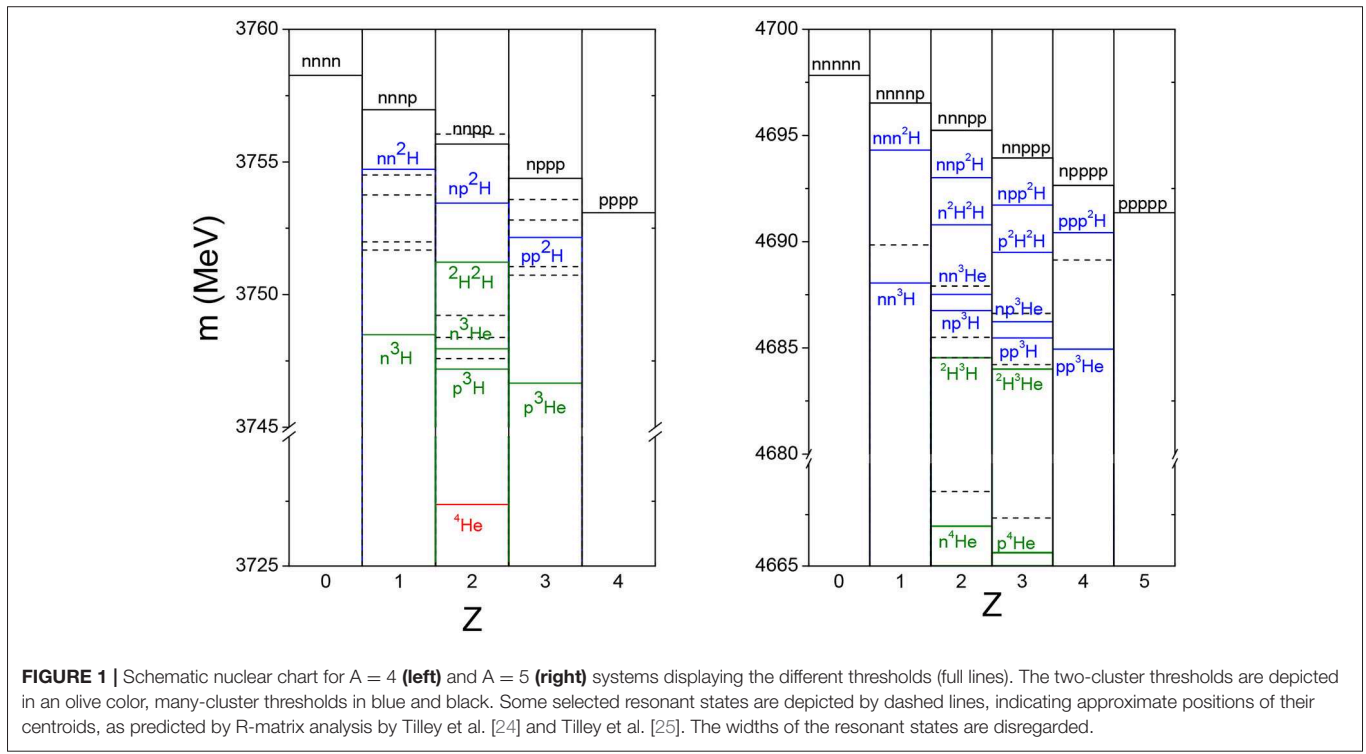

**FIGURE 1 |** Schematic nuclear chart for A = 4 **(left)** and A = 5 **(right)** systems displaying the different thresholds (full lines). The two-cluster thresholds are depicted in an olive color, many-cluster thresholds in blue and black. Some selected resonant states are depicted by dashed lines, indicating approximate positions of their centroids, as predicted by R-matrix analysis by Tilley et al. [24] and Tilley et al. [25]. The widths of the resonant states are disregarded.

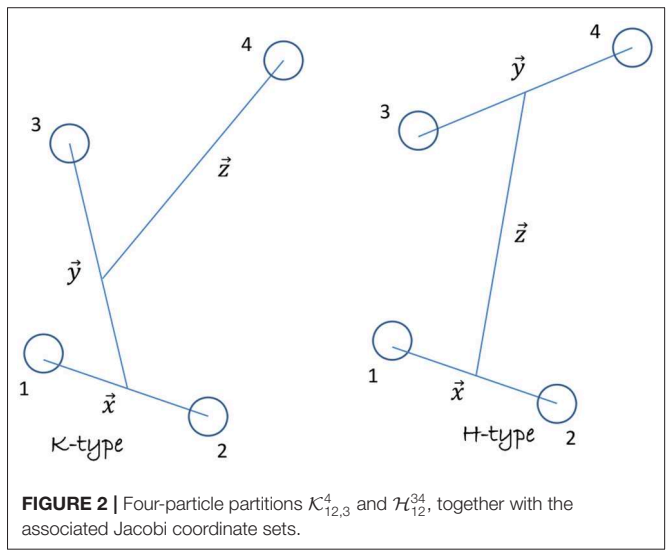

**FIGURE 2 |** Four-particle partitions $\mathcal{K}^4_{12,3}$ and $\mathcal{H}^{34}_{12}$, together with the associated Jacobi coordinate sets.

of the components $\mathcal{K}^l_{ij,k}$ and $\mathcal{H}^{kl}_{ij}$ incorporate all the possible 3+1 and the 2+2 particle channels, respectively, as illustrated in **Figure 2**. Here, we are interested in nuclear problems, involving protons and neutrons. Within the isospin formalism, neutrons and protons are treated as isospin-degenerate states of the same particle: the nucleon. Then, the FY components, which differ by the order of the particle indexing, are related due to the symmetry of particle permutation. There remain only two independent FYCs, which are further denoted $\mathcal{K} \equiv \mathcal{K}^4_{12,3}$ and $\mathcal{H} \equiv \mathcal{H}^{34}_{12}$ by omitting their particle indexes. For FY equations for

a case of four identical particles (see [11, 43]):

$$(E - H_0 - V_{12})\mathcal{K} = V_{12}(P^+ + P^-)\left[(1 + Q)\mathcal{K} + \mathcal{H}\right],$$
$$(E - H_0 - V_{12})\mathcal{H} = V_{12}\tilde{P}\left[(1 + Q)\mathcal{K} + \mathcal{H}\right], \quad (3)$$

Each FY component $\mathcal{F} = (\mathcal{K}, \mathcal{H})$ has its natural expression in its proper set of Jacobi coordinates, as depicted in **Figure 2**. However, they may be as well-considered as a function of any set of Jacobi coordinates and converted for one coordinate set into another one by using the particle permutation operators, which are summarized as follows:

$$P^+ = (P^-)^{-1} \equiv P_{23}P_{12},$$
$$Q \equiv -P_{34},$$
$$\tilde{P} \equiv P_{13}P_{24} = P_{24}P_{13},$$

where $P_{ij}$ indicates the operator permuting particles $i$ and $j$.

In terms of the FYCs, the total wave function of an $A = 4$ system is given by:

$$\Psi = \left[1 + (1 + P^+ + P^-)Q\right](1 + P^+ + P^-)\mathcal{K}$$
$$+ (1 + P^+ + P^-)(1 + \tilde{P})\mathcal{H}. \quad (4)$$

Each FY component $\mathcal{F} = (\mathcal{K}, \mathcal{H})$ is considered as a function, described in its proper set of Jacobi coordinates, as depicted in **Figure 2**.

The angular, spin, and isospin dependence of these components are described using the tripolar harmonics $\mathcal{Y}_\alpha(\hat{x}, \hat{y}, \hat{z})$, i.e.,:

$$\langle \vec{x}\vec{y}\vec{z}|\mathcal{F}\rangle = \sum_\alpha \frac{\mathcal{F}_\alpha(xyz)}{xyz} \mathcal{Y}_\alpha(\hat{x}, \hat{y}, \hat{z}). \quad (5)$$

The quantities $\mathcal{F}_\alpha(xyz)$ are called the regularized radial FY amplitudes, where the label $\alpha$ holds for a set of 10 intermediate quantum numbers describing a given four-nucleon quantum state ($J^\pi$, $T$, $\mathcal{T}_z$). By using the LS-coupling scheme, the tripolar harmonics are defined for components of K and H type, respectively, by

$$\mathcal{Y}_{\alpha_K} \equiv \left\{ \left[ \left[ (l_x l_y)_{l_{xy}} l_z \right]_L \left[ \left( (s_1 s_2)_{s_x} s_3 \right)_{S_3} s_4 \right]_S \right]_{J^\pi} \mathcal{M} \right.$$
$$\left. \otimes \left[ \left( (t_1 t_2)_{t_x} t_3 \right)_{T_3} t_4 \right]_{T \mathcal{T}_z} , \right. \tag{6}$$

$$\mathcal{Y}_{\alpha_H} \equiv \left\{ \left[ \left[ (l_x l_y)_{l_{xy}} l_z \right]_L \left[ (s_1 s_2)_{s_x} (s_3 s_4)_{s_y} \right]_S \right]_{J^\pi} \mathcal{M} \right.$$
$$\left. \otimes \left[ t_1 t_2 \right)_{t_x} (t_3 t_4)_{t_y} \right]_{T \mathcal{T}_z} . \tag{7}$$

The FY equations were originally derived to treat systems of particles interacting by pairwise short-range interactions. Nevertheless, these equations can be modified with relative ease to include three-body forces (3BF). This has been achieved for the first time in the work of the Bochum group [44]. In implementing three-nucleon forces, we have followed quite a similar but nevertheless slightly optimized strategy [43]. It is worth noticing a recent work by Kamada [45] presenting a systematic derivation of the four-body FY equations by including three-body forces.

## 2.2. Treatment of the Coulomb Interaction

One of the more delicate issues in solving few-particle scattering problems is the proper treatment of the Coulomb interaction. Due to the long-range nature of Coulomb potential in coordinate space, or, equivalently, due to its singular behavior in momentum space, the standard approach of the scattering theory based on expansion in free waves in the asymptote region is not appropriate. Indeed, the FY equations, as presented in the previous section, are formulated for short-range interactions and are not appropriate for handling scattering problems including Coulomb interaction.

For a three-body system, the proper mathematical formalism to include Coulomb interactions was proposed by Merkuriev [46]. This formalism is valid both for attractive and for repulsive Coulomb forces. The problem becomes considerably simpler if only one repulsive Coulomb interaction is present (only two out of three particles are charged by equal sign charges) like for a proton-deuteron scattering. For this particular case, there are several alternative prescriptions to handle Coulomb interaction force [47–52]. They are based on inserting, fully or partially, the Coulomb potential $V_C(x,y)$, in the left-hand side of the Faddeev equation

$$[E - H_0 - V(x) - V_C(x,y)]\Psi(x,y) = (V(x) - V_c^l(x,y))(P^+ + P^-)\Psi(x,y). \tag{8}$$

In this way, the long-range part of the Coulomb interaction $V_c^l(x, y)$ is subtracted on the right-hand side of the Faddeev equation and is appropriately compensated by the term $V_C(x,y)$ on the left-hand side, thus accounting for the Coulomb asymptotic wave function in the scattering channel.

On the contrary, for an N>3 case, no such modifications of FY equations existed prior to our work. Only in the work of Filikhin and Yakovlev [53] has this problem been partly addressed by being limited to S-wave approximation.

### 2.2.1. Formulation, à la Merkuriev [46]

In this work, we present two alternatives for how to treat Coulomb interaction for N>3 systems: following the strategy of Merkuriev, and following the method proposed by Sasakawa and Sawada [48]. It is worth mentioning that many alternative treatments of the repulsive Coulomb interaction, corresponding to the three-body approaches of Noble [47] and Chen et al. [50], may be formally spanned under Merkuriev's approach by considering different forms of splitting the Coulomb interaction into short- and long-range parts.

In this section, we propose a generalization of the four-body FY equations following Merkurievs approach to the three-body system [46]. We start by splitting the Coulomb potential $V_{ij}^C$ into two parts: short-range $V_{ij}^s$ and long-range $V_{ij}^l$, such that $V_{ij}^C = V_{ij}^s + V_{ij}^l$. This is realized by means of appropriate cut-off functions $\chi^{s_a}(x_{ij}, y_a, z_a)$, depending on the radial parts of Jacobi coordinates as they are depicted in **Figure 2**:

$$V_{ij}^{s_a} = \chi_a^s(x_{ij}, y_a, z_a)V^C(x_{ij}). \tag{9}$$

We introduce three different forms of splitting

$$V_{ij}^C(x_{ij}) = V_{ij}^{l_K}(x_{ij}, y_{ij,k}) + V_{ij}^{s_K}(x_{ij}, y_{ij,k}), \tag{10}$$

$$V_{ij}^C(x_{ij}) = V_{ij}^{l_H}(x_{ij}, y_{lk}) + V_{ij}^{s_H}(x_{ij}, y_{lk}), \tag{11}$$

$$V_{ij}^C(x_{ij}) = V_{ij}^{l_\rho}(x_{ij}, \rho_{ij}) + V_{ij}^{s_\rho}(x_{ij}, \rho_{ij}), \tag{12}$$

and, following the steps leading to three-body Merkuriev equations, we reformulate the four-body equations as follows:

$$(E - H_0 - V_{12} - V_{13}^{l_K} - V_{23}^{l_K} - V_{14}^{l_\rho} - V_{24}^{l_\rho} - V_{34}^{l_\rho})\mathcal{K}_{12,3}^4$$
$$= V_{12}^{s_K}(\mathcal{K}_{23,1}^4 + \mathcal{K}_{31,2}^4) + V_{12}^C(\mathcal{K}_{23,4}^1 + \mathcal{K}_{31,4}^2 + \mathcal{H}_{23}^{14} + \mathcal{H}_{31}^{24}), \tag{13}$$

$$(E - H_0 - V_{12} - V_{34}^{l_H} - V_{13}^{l_\rho} - V_{23}^{l_\rho} - V_{14}^{l_\rho} - V_{24}^{l_\rho})\mathcal{H}_{12}^{34}$$
$$= V_{12}^{s_H}\mathcal{H}_{12}^{34} + V_{12}^{s_\rho}(\mathcal{K}_{34,1}^2 + \mathcal{K}_{34,2}^1). \tag{14}$$

One may easily verify that, by summing these equations, one obtains Schrödingers equation for the total wave function of the system : $\Psi = \sum \mathcal{K}_{ij,k}^l + \sum \mathcal{H}_{ij}^{lk}$. On the other hand, the asymptotes of the different binary channels become perfectly separated. To demonstrate this feature, let us investigate the FY component $\mathcal{K}_{12,3}^4$ associated with Equation (13). This component is meant to incorporate the asymptote of the (123)+4 particle channel and is directly coupled with the components $\mathcal{K}_{23,4}^1$, $\mathcal{K}_{31,4}^2$, $\mathcal{H}_{23}^{14}$, and $\mathcal{H}_{31}^{24}$, which are not proper to the (123)+4 particle channel. Nevertheless, in Equation (13) this coupling is ensured only by the short-range interaction $V_{12}^{s_K}$. It remains coupled with the components $\mathcal{K}_{23,1}^4$ and $\mathcal{K}_{31,2}^4$ by the long-range interaction terms, but all these components belong to the same (123)+4 particle channel. Very similar behavior is preserved by the component $\mathcal{H}_{12}^{34}$, associated with Equation (14). This component contributes to the asymptote of the (12)+(34) particle channel and is coupled by a long-range interaction term only with the component $\mathcal{H}_{34}^{12}$ belonging to the same binary channel. Therefore, the modified FY Equations (13)-(14) uncouple the asymptotes belonging to different binary scattering channels even when the Coulomb interaction is present. Their uncoupling properties are in this way similar to the original FY equations involving only short-range interactions.

### 2.2.2. Alternative Formulation, à la Sasakawa and Sawada [48]

For each FYC, one introduces an auxiliary long-range potential in their asymptote describing an effective Coulomb repulsion between the fragments of the associated binary channel $V_{z_\alpha}^l(z_\alpha)$, such that:

$$V_{z_\alpha}^l(z_\alpha \to \infty) = \frac{C_\alpha}{z_\alpha}, \tag{15}$$

where, for a type-K component, defined by the particle ordering $\alpha = \binom{4}{12,3}$ :

$$C_{12,3}^4 \equiv C_4 = \sqrt{\frac{2m_4(M - m_4)}{Mm}}; \tag{16}$$

and for type-H component with $\alpha = \binom{34}{12}$,

$$C_{12}^{34} \equiv C_{34}^{12} = \sqrt{\frac{2(m_1 + m_2)(m_3 + m_4)}{Mm}}. \tag{17}$$

We denote by $m_i$ the mass of nucleon $i$, and by $M = \sum_i m_i$, the total mass of the system. FY equations are reformulated by subtracting this long-range potential in their left-hand side. These auxiliary potential terms are compensated by introducing appropriate terms in the right-hand side of FY equations:

$$(E - H_0 - V_{12} - q_4(q_1 + q_2 + q_3)V_{z_4}^l)\mathcal{K}_{12,3}^4$$
$$= V_{12}(\mathcal{K}_{23,1}^4 + \mathcal{K}_{31,2}^4 + \mathcal{K}_{23,4}^1 + \mathcal{K}_{23,4}^2 + \mathcal{H}_{23,4}^{14} + \mathcal{H}_{31,4}^{24}),$$
$$- q_1q_2(V_{z_1}^l\mathcal{K}_{23,4}^1 + V_{z_2}^l\mathcal{K}_{31,4}^2 + V_{z_{23}}^l\mathcal{H}_{23}^{14} + V_{z_{13}}^l\mathcal{H}_{13}^{24}) \tag{18}$$

$$(E - H_0 - V_{12} - (q_1 + q_2)(q_3 + q_4)V_{z_{12,3}^l}^l)\mathcal{H}_{12}^{34}$$
$$= V_{12}(\mathcal{K}_{34}^{12} + \mathcal{K}_{34,1}^1 + \mathcal{K}_{34,2}^1) - q_1q_2(V_{z_1}^l\mathcal{K}_{34,1}^1 + V_{z_1}^l\mathcal{K}_{34,2}^1), \tag{19}$$

where $q_i$ is the charge of particle $i$. The auxiliary potential terms $V_{z_\alpha}^l(z_\alpha)$ in the left- and right-hand sides of equations are balanced in such a way that they compensate each other once all 18 FY equations are added to recover Schrödingers equation.

Further, we are interested in uncoupling of the wave components describing different two-cluster scattering channels. To see how well these components uncouple in their asymptotes, let us analyze the first equation associated with a component $\mathcal{K}_{12,3}^4$. In the right-hand side of this equation, components $\mathcal{K}_{23,4}^1$, $\mathcal{K}_{31,4}^2$, $\mathcal{H}_{23}^{14}$, and $\mathcal{H}_{13}^{24}$ are present, which are not proper to the $(123) + 4$ elastic channel of the component $\mathcal{K}_{12,3}^4$. As an example, component $\mathcal{K}_{23,4}^1$, associated with the $(234) + 1$

scattering channel, is coupled with the $\mathcal{K}_{12,3}^4$ in this equation by the potential term

$$V_{12} - q_1q_2V_{z_1}^l = q_1q_2 \left( \frac{1}{x_{12}} - \frac{\sqrt{\frac{2m_1(M - m_1)}{Mm}}}{z_1} \right), \tag{20}$$

which behaves as $O((z_1)^{-3})$ in the $z_1 >> \max(x_{23}, y_{23}^4)$ region, which defines the asymptote of the $(234) + 1$ scattering channel. One may reach the same conclusion relative to the coupling between the components of types $\mathcal{K}$ and $\mathcal{H}$. Thus, the asymptotic coupling between the components belonging to different binary channels is realized by the effective potential terms decaying as $O((z_1)^{-3})$ and thus is strongly suppressed relative to the original Coulomb potential. One should mention, however, that such uncoupling is not ensured for the case when breakup in three (or four) clusters is energetically allowed.

By comparing the approach of Equations (13)–(14) to the one following Equations (18)–(19), one may readily conclude that, in the first approach, the FY components are more properly uncoupled by the exponentially decaying potential terms. Nevertheless, the second formalism requires less effort implement numerically. In the following section, we will present some results demonstrating that these two approaches work equally well for the nuclear problem, where only repulsive Coulomb interactions are present.

## 3. RESULTS

### 3.1. Models

The results presented in this study are obtained using realistic nuclear Hamiltonians. The realistic nucleon-nucleon (NN) potentials contain several adjustable parameters, which are tuned in order to reproduce experimental NN scattering data and the properties of a deuteron with very high accuracy. Three different NN potentials, Argonne $v_{18}$ (AV18) [54], INOY04 [55], and Idaho N3LO (I-N3LO) [56], are used in this work. The AV18 model is a phenomenological potential, which is defined in configuration space and is local. The longest-range part of the AV18 potential is determined by one-pion exchange and electromagnetic NN interaction terms, but its short-range part is fully phenomenological.

The locality of the NN force assumed in the pioneering high-accuracy NN interaction models was due to numerical convenience. Nevertheless, it was soon realized that such models suffer from the underbinding problem when describing $A > 2$ nuclei [57–59]. The inclusion of non-local interaction terms allows the off-shell structure of the potential to change and may strongly affect the description of the $A > 2$ sector. This feature has been explored by Doleschall [55, 60, 61], who constructed a set of phenomenological potentials. The internal parts of these potentials are built by employing highly non-local form factors (the INOY04 model non-locality range is approximately 2 fm), whereas their outside parts are local and are defined by the Yukawa potential representing one-pion exchange. The NN interaction models of Doleschall et al., and in particular INOY04, were able to overcome the lack of binding energy in

**TABLE 1 |** Parameters of the local N2LO 3BF employed in this work.

| Λ (MeV) | $c_D$ | $c_E$ | $c_1$ | $c_3$ | $c_4$ | References |
|---|---|---|---|---|---|---|
| 350 | −0.2 | 0.205 | −0.81 | −3.2 | 5.4 | [72] |
| 400 | −0.2 | 0.098 | −0.81 | −3.2 | 5.4 | [72] |
| 450 | −0.2 | −0.016 | −0.81 | −3.2 | 5.4 | [72] |
| 500 | −0.0411 | 0.945 | −0.81 | −3.2 | 5.4 | [71] |
| 500 | −0.2 | −0.205 | −0.81 | −3.2 | 5.4 | [70] |

*In the last column, a reference to the original work where the values of the parameters $c_D$ and $c_E$ were established is provided.*

**TABLE 2 |** Comparison of the p-$^3$He singlet ($J^\pi = 0$) and triplet ($J^\pi = 1$) scattering lengths calculated by the approach of Equations (13)–(14) to those obtained by solving Equations (18)–(19).

| $J^\pi$ | $0^+$ | $1^+$ |
|---|---|---|
| Equations (13)–(14) | 11.92 | 9.346 |
| Equations (18)–(19) | 11.86 | 9.302 |

*Scattering lengths are provided in units of fm.*

the three-nucleon sector, namely $^3$H and $^3$He, without explicitly using three-nucleon forces and still accurately reproduce NN observables [11, 55].

From the early 2000s, inspired by the works of Weinberg [62], a new generation of nuclear forces appeared based on chiral effective field theory [63, 64]. Chiral effective field theory provides a powerful framework with which to link the NN potentials with the pion-nucleon ones but, at the same time, construct systematically, order by order, an improvable scheme to build consistent multinucleon forces as well as control the uncertainties in their determination. The chiral NN interaction model developed up to next-to-next-to-next-to-leading order by the Idaho group [56], denoted here as I-N3LO, remains one of the most successful descriptions of the NN interaction.

Realistic nucleon-nucleon interaction models are nowadays able to describe all the available scattering data in the two nucleon sector almost perfectly. Studies of heavier nuclei are therefore required in order to test and validate these interaction models. However, calculations of the trinucleon binding energies already reveal an underbinding problem: most of the nucleon-nucleon potentials fail to reproduce binding energies of triton and $^3$He. A single exception is provided by the INOY potentials, which, employing non-local form factors, are adjusted at NN level to also reproduce the binding energy of $^3$H. Nevertheless, these models turn out to be too soft, compressing, and overbinding $^4$He [11, 65], leading to high saturation density of the nuclear matter [66] as well as severe overbinding of heavier nuclei [67]. The natural remedy is the introduction of three-nucleon forces, which appear in any theoretically motivated nuclear interaction model. It should be noted that only models based on effective-field theory provide a systematic hierarchy between two-nucleon and multi-nucleon forces. Regardless of the three-nucleon interaction model, these forces have some adjustable parameters.

There are several different three-nucleon force models that can be used in conjunction with AV18 and the chiral effective field potentials of Epelbaum et al. [63] or Machleidt and Entem [64]. Notably, with AV18 NN potential, we will employ the Urbanna IX (UIX) three-nucleon interaction model of Pudliner et al. [68], adjusted in order to improve description of the three-nucleon binding energies as well as the nuclear matter saturation density.

In Navrátil [69], a three-nucleon force employing local momentum-space regulators, and developed up to next-to-next-to-leading order was proposed. In Gazit et al. [70], two unknown coupling constants, $c_D$ and $c_E$, of this 3BF were adjusted to reproduce triton binding energy and $\beta$-decay half-life simultaneously. In a recent work [71], it was found that the relation between the low-energy constants (LECs) $c_D$ and $c_E$, determining the three-nucleon contact interaction and the two-nucleon contact axial current, was given erroneously. A new parametrization of the last force was also provided. In our work, we will essentially use the last parametrization of the force of Marcucci et al. [71] using cutoff $\Lambda = 500$ MeV (see **Table 1**). It is worth mentioning that the two parameterizations, one of Gazit et al. [70] and one of Marcucci et al. [71], provide almost identical predictions for all the nuclear observables considered here.

An alternative strategy to fix $c_D$ and $c_E$ coupling constants was followed by Roth et al. [72]. Those authors noticed that heavier nuclei are overbound when 3BF of Gazit et al. [70] is used. A new set of three-nucleon forces were proposed using lower cutoff values, $\Lambda = 350$, 400, and 450 MeV, which describe the binding energies of the medium mass nuclei along the dripline better.

For convenience, the different parameterizations of 3BF used in what follows in conjunction with I-N3LO NN interaction will be referred to by the cutoff value $\Lambda$ regularizing this force. In **Table 1**, we provide the parameters of the different 3BFs tested in this work.

## 3.2. Coulomb Phaseshifts

As described in section 2.2, the implementation of the Coulomb interactions represents a real challenge for the few-nucleon scattering problem. Two different methods have been proposed to implement the Coulomb force in nucleon-trinucleon scattering. We present in **Tables 2, 3** a comparison among these two approaches. One may see that, regardless of the fact that the method based on Equations (13)–(14) is formally more appropriate, in Practice, the two methods provide almost identical results. Even for p-$^3$He scattering length calculations, where the effect of Coulomb repulsion should be the most appreciable, the two methods provide indistinguishable results within the numerical accuracy. A small discrepancy might be still observed in calculating negative parity phaseshifts (see **Table 3** corresponding to $E_p = 2.25$ MeV), which might be related to the importance of the triton polarizations terms, while these terms are partly screened in the approach based on Equations (18)–(19).

## 3.3. Description of the Four-Nucleon Scattering

The main goal of theoretical nuclear physics is to construct a reliable model describing the nuclear structure and reactions.

**TABLE 3** | The same as in **Table 2** but for the scattering phase shifts ($^{2s+1}L_J$) and mixing angles ($\epsilon^{-J^{\pi}}$) at $E_p = 2.25$ MeV, both presented in degrees.

| | $^1S_0$ | $^3S_1$ | $^3D_1$ | $\epsilon^{1+}$ | $^3P_0$ | $^1P_1$ | $^3P_1$ | $\epsilon^{1-}$ | $^3P_2$ |
|---|---|---|---|---|---|---|---|---|---|
| Equations (13)–(14) | −41.57 | −35.49 | −0.28 | −0.58 | 7.74 | 17.75 | 10.84 | 8.43 | 16.41 |
| Equations (18)–(19) | −41.50 | −35.42 | −0.28 | −0.57 | 8.15 | 17.72 | 11.25 | 8.55 | 16.66 |

Realistic nucleon-nucleon interaction models are built to reproduce available data in two-nucleon sectors. In addition, three-nucleon forces are usually introduced and adjusted to reproduce the ground state binding energies of triton ($^3$H), $^3$He, and $^4$He. Nevertheless, the binding energies of the stable nuclei, appearing along the dripline, are strongly correlated and thus provide only limited insight into nuclear forces. Scattering experiments, allowing unbound structures far from the stability to be accessed, remain the richest tool to study the properties of the underlying nuclear interaction.

Three-nucleon systems have been extensively studied throughout the last two decades [73]. Realistic nuclear Hamiltonians provide a satisfactory description of the uttermost part of the trinucleon data. Still, there remain some discrepancies, like in the description of the analyzing powers ($A_y$-puzzle) and some breakup observables (space-star anomaly) [74, 75], which have not yet been addressed by any NN+3BF model. It is noteworthy that these problematic observables are relatively small, representing only a few percentiles of the total scattering cross-section.

Three-nucleon systems remain relatively simple, due to the absence of any thresholds (apart from the three-particle breakup one) or resonant structures in the continuum. The two experimentally accessible systems, $^3$H and $^3$He, are mirror systems and thus exhibit very similar properties. Four- and five-nucleon systems, accommodating several resonant states and a rich threshold structure in the continuum, therefore present interesting theoretical laboratories for testing the nuclear interactions.

The elastic neutron scattering on $^3$H, being a process free from Coulomb interaction, is the simplest four-nucleon reaction to describe theoretically. Unfortunately, due to nuclear safety regulations, experiments with tritium are scarce. Nevertheless, some successful measurements were realized in the 1970s. In particular, very accurate measurement of the total neutron-tritium cross-section was realized by Phillips et al. [76].

In **Figure 3**, we compare our calculated results with the ones of this measurement. There are two important energy regions for the elastic neutron scattering on $^3$H: the zero energy region (S-waves) and the region of P-wave resonances.

At very low energies, the process is dominated by neutron scattering in S-waves relative to the target. These waves are governed by the Pauli repulsion between the neutron projectile and those present in the tritium target. Due to this repulsion, the scattering process is mostly peripheral, and therefore the calculated scattering lengths strongly correlate with the size of the target nucleus and consequently with the predicted tritium binding energy. Thus, the nuclear interaction models that tend to underbind the triton overestimate the n-$^3$H cross-section at low energy. By adjusting the triton binding energy, either by means of three-nucleon force or by the presence of non-locality in NN interaction (INOY models), the agreement with the experimental n-$^3$H cross-section significantly improves in the zero-energy limit. It is worth noticing that some minor differences still remain between the models, i.e., the predictions of INOY04 or I-N3LO+3BF($\Lambda = 500$ MeV) agree with a lower bound of the zero-energy cross-section, whereas AV18+UIX agrees with an upper one. These differences could be resolved by comparing the calculated spin-dependent ($a_0$ and $a_1$) scattering lengths. Unfortunately, there is quite a large discrepancy between the measured coherent scattering lengths $a_c$, defined as $a_c = \frac{a_0}{4} + \frac{3a_1}{4}$, and the inferred spin-dependent values. The measurement of the coherent scattering length $a_c$ constrains the values of $a_1$ and $a_0$ to a linear band, while the measurement of the total n-$^3$H cross-section constrains them in an elliptic one. The spin-dependent values $a_i$ result from the intersection of these two bands, but their practical determination is not free of ambiguities due to experimental errors. This is illustrated in **Figure 4**, together with the predictions of the nuclear models considered.

The total neutron-triton cross-section peaks at around 3 MeV. This peak results from the interference of four broad negative parity resonant states present in the $^4$H nucleus. The accurate description of the n-$^3$H cross-section in this resonance region turns out to be a very challenging problem for nuclear interaction models. Most of these models fail to provide sufficient attraction for negative parity states (essentially P-waves), providing a very flat structure. In this context, the role of 3BF is quite essential and far from trivial. First, by adding a 3BF that reproduces the triton binding energy, one automatically reduces the contribution of the partial cross-sections in the positive parity states $J^{\pi} = 0^+$ and $1^+$. Then, the required increase in negative parity cross-section should fill the existing gap in the resonance region and compensate for the reduction from the positive parity state contribution.

Among the models described in **Figure 3**, UIX fails to boost the contribution from the negative parity states in the total cross-section. Therefore, the net effect is a reduction of the total cross-section in the resonance region. Of the three NN interaction models considered, I-N3LO provides the most attraction in the negative parity states. The three-nucleon interaction model with a cutoff $\Lambda = 500$ MeV further improves the agreement between the calculated and measured cross-sections, describing almost ideally the experimental data of Phillips et al. [76]. Notice, however, that the parameterizations of the same 3BF employing lower cutoff values from Roth et al. [72] are not successful, underpredicting the total cross-section. It is also worth noting that the discrepancy in the resonance region is increased by reducing the value of the cutoff $\Lambda$. Very similar consequences are observed when calculating the binding energies of the P-shell nuclei [72].

The proton scattering on $^3$He is a nuclear mirror process to neutron scattering on $^3$H. The presence of Coulomb interaction makes the proton scattering on the $^3$He cross-section diverge at small angles, so one is not able to study the cross-section of this process with the same ease as for the n-$^3$H case. Nevertheless, experimental differential cross-sections are much more abundant for the p-$^3$He case since they are easier to measure.

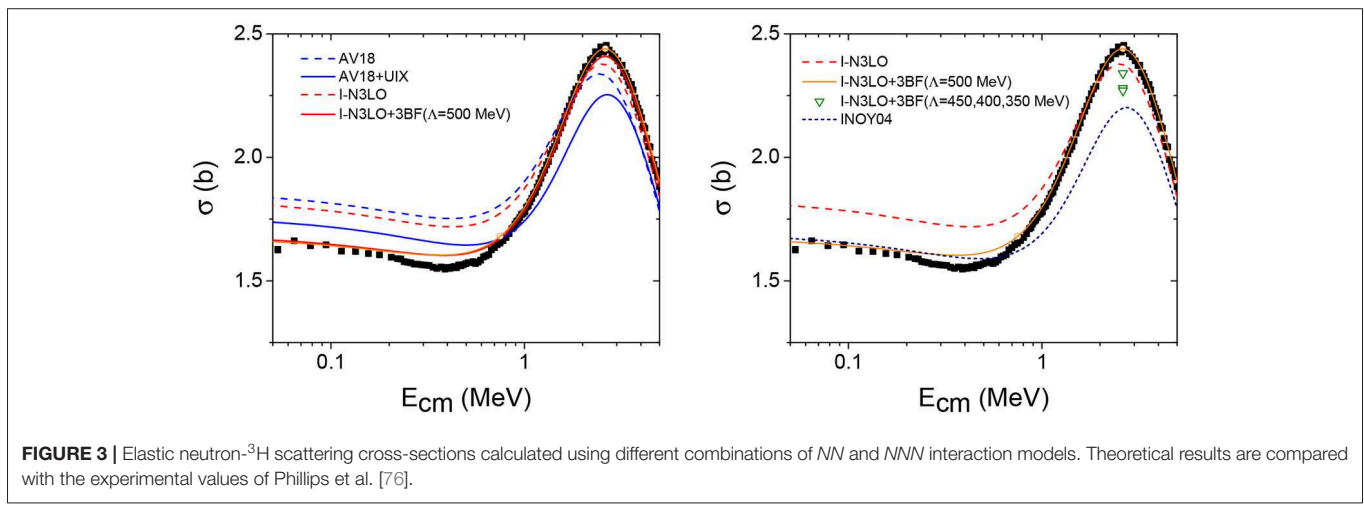

**FIGURE 3 |** Elastic neutron-$^3$H scattering cross-sections calculated using different combinations of *NN* and *NNN* interaction models. Theoretical results are compared with the experimental values of Phillips et al. [76].

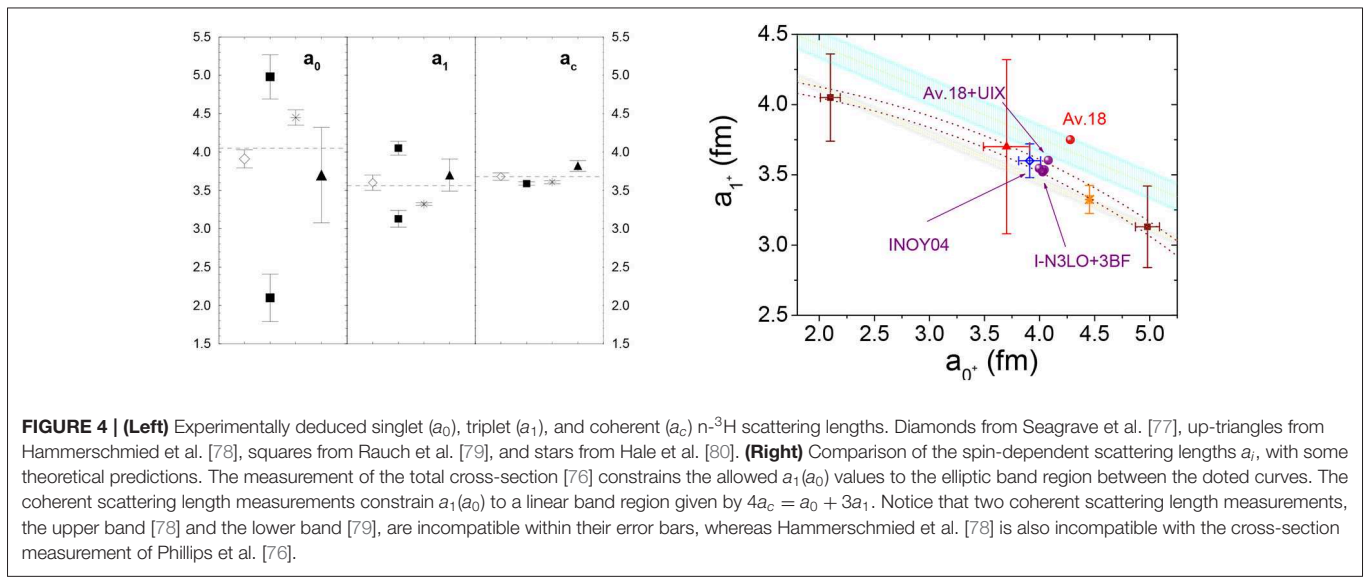

**FIGURE 4 | (Left)** Experimentally deduced singlet ($a_0$), triplet ($a_1$), and coherent ($a_c$) n-$^3$H scattering lengths. Diamonds from Seagrave et al. [77], up-triangles from Hammerschmied et al. [78], squares from Rauch et al. [79], and stars from Hale et al. [80]. **(Right)** Comparison of the spin-dependent scattering lengths $a_i$, with some theoretical predictions. The measurement of the total cross-section [76] constrains the allowed $a_1(a_0)$ values to the elliptic band region between the dotted curves. The coherent scattering length measurements constrain $a_1(a_0)$ to a linear band region given by $4a_c = a_0 + 3a_1$. Notice that two coherent scattering length measurements, the upper band [78] and the lower band [79], are incompatible within their error bars, whereas Hammerschmied et al. [78] is also incompatible with the cross-section measurement of Phillips et al. [76].

We have displayed in **Figure 5** the calculated p-$^3$He scattering observables for an incident energy of 4.05 MeV. The same observables for protons of 5.54 MeV are displayed in **Figure 6**. Calculated results arrived at by employing Equations (13)–(14) to take Coulomb into account are compared with the available experimental data from McDonald et al. [81], Alley and Knutson [82], Fisher et al. [83], and Daniels et al. [84]. The energy region considered is still marked by the important contribution of the negative parity $^4$Li resonances. Notably, due to the presence of the repulsive Coulomb interaction, these resonances manifest at slightly higher energies for the p-$^3$He case relative to the n-$^3$H one. In **Figure 5**, relevant for 4.05 MeV protons, the theoretical values corresponding to all the aforementioned nuclear interaction models are displayed.

By studying the angular differential cross-section, one may observe quite similar properties as previously outlined for the n-$^3$H total cross-section at the resonance peak. The I-N3LO NN interaction model provides the most accurate description of the data if used in conjunction with a 3BF with a cut-off parameter of $\Lambda = 500$ MeV. Other models tend to underestimate the scattering cross-section, while the net differences are quite small. The most relevant observable for studying the model dependence remains the analyzing power $A_{y0}$. At the maximum of $A_{y0}$, one observes an up to 30% spread between the different model predictions. Once again, I-N3LO+3BF($\Lambda = 500$ MeV) turns to be the most successful in describing experimental data and sits almost on top of it. Nevertheless, the deviation relative to the experimental data is approximately 2% of the absolute cross-section values, which is comparable to the discrepancy present in the three-nucleon sector (Ay-puzzle). The analyzing powers are simply much weaker, in absolute values, in nucleon-deuteron scattering, and therefore discrepancies seem to be much more substantial.

When comparing the effect of different the 3BFs employed in conjunction with I-N3LO NN interaction in a similar way as outlined for the n-$^3$H cross-section case, the description of p-$^3$He deteriorates when the cutoff $\Lambda$ is reduced from its original value, 500 MeV. This feature does not seem to be related to the importance of reproducing the tritium beta decay half-life, as only the Marcucci et al. [71] model accounts for it. The 3BF model of Gazit et al. [70] provides almost identical results to those obtained using the 3BF from Marcucci et al. [71], both using the value $\Lambda = 500$ MeV. Another quite straightforward answer would be the importance of maintaining consistency between the regulators in NN and three-nucleon interaction. Nevertheless, while the I-N3LO interaction is regulated by employing the same cutoff value of $\Lambda = 500$ MeV, the expressions of these regulators are quite different for NN and 3BF.

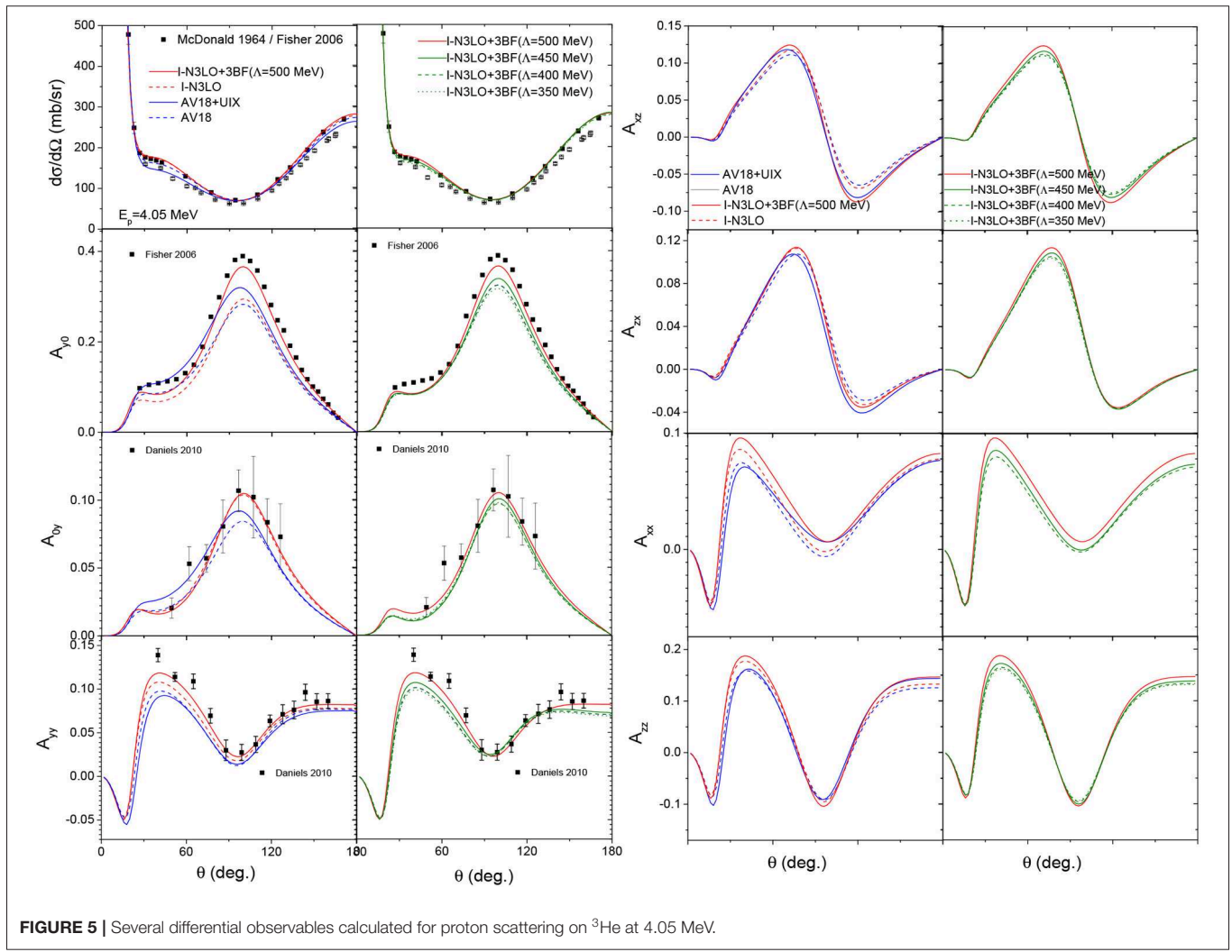

**FIGURE 5 |** Several differential observables calculated for proton scattering on $^3$He at 4.05 MeV.

In order to consider scattering at even higher energies—for $E_n \gtrsim 8$ MeV neutrons or $E_p \gtrsim 7$ MeV protons—one should take the presence of the three- (or/and even four-) particle breakup channels into account. The description of such processes is far beyond the reach of the standard techniques based on imposing proper boundary conditions in configuration space (or treating singularities in a multidimensional kernel of integral equations formulated in momentum space). Nevertheless, one may avoid these complications by employing complex scaling or complex energy methods, as has been successfully demonstrated in Carbonell et al. [85]. The scattering in n-$^3$H and p-$^3$He systems has been accurately described above d+N+N and also above 4N thresholds in recent work [17, 86, 87]. In particular, it has been found that description of the analyzing power improves in these systems once energy is increased. This fact is clearly demonstrated in **Figure 7**. The interplay of the 3BF has not yet been explored in studying four-nucleon scattering above the three-particle breakup threshold. Nevertheless, some indications are present that the calculated total elastic and breakup cross-sections correlate with the predicted binding energy of the target nucleus, as illustrated in **Figure 8**. This feature is attributed to the importance of correctly positioning the thresholds in describing low-energy scattering cross-sections. In the vicinity of a threshold, and due to the kinematic form factor, the breakup cross-section increases with the available kinetic energy. Conversely, the elastic cross-sections tend to decrease with energy. For the models reproducing tri-nucleon binding energies properly, the obtained n-$^3$H and p-$^3$He cross-sections are successfully described in the intermediate energy region.

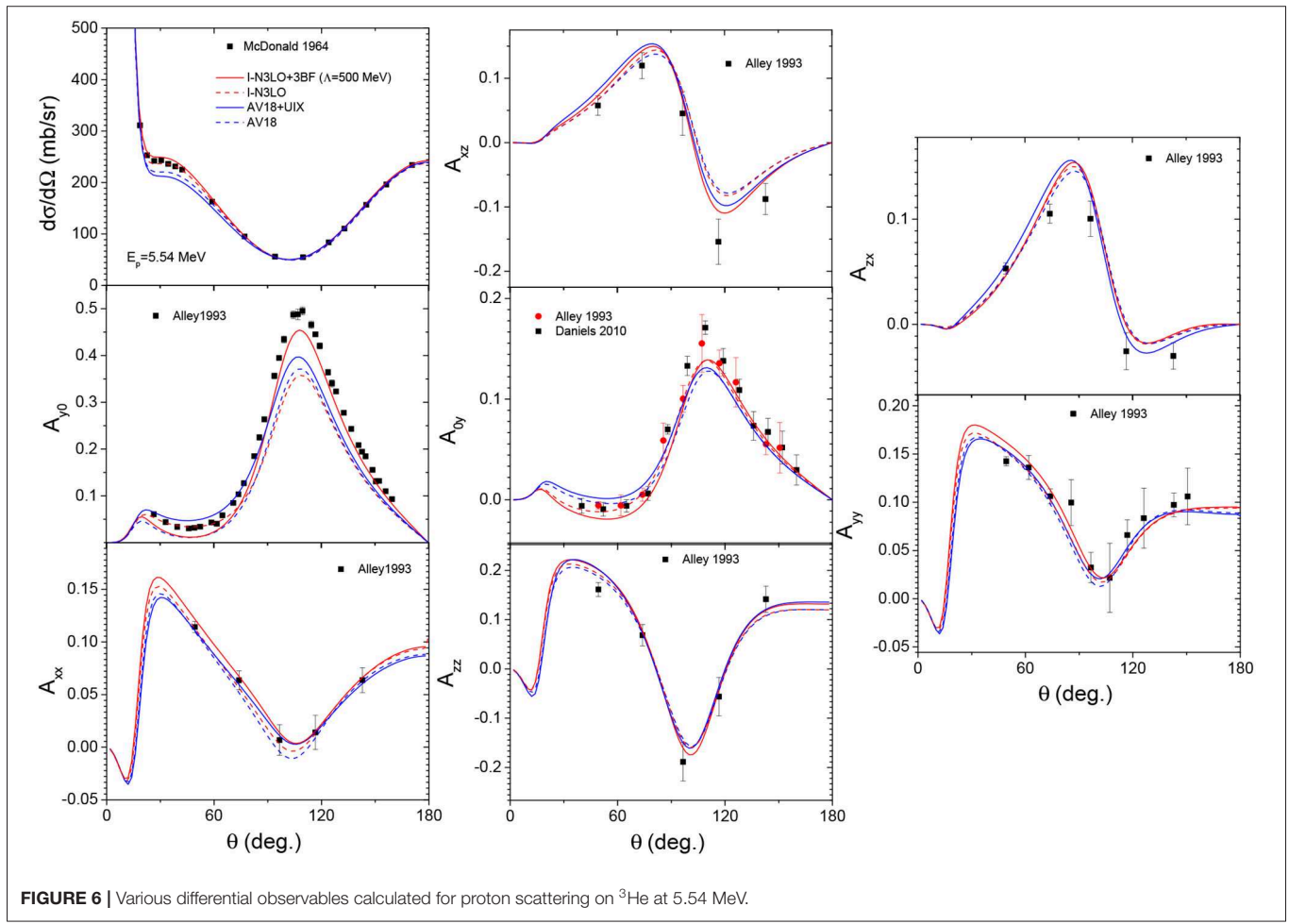

**FIGURE 6 |** Various differential observables calculated for proton scattering on $^3$He at 5.54 MeV.

When considering elastic differential cross-sections, some discrepancies have been found when studying 22.1 MeV neutron scattering on $^3$H, in particular at the cross-section minima (see **Figure 7**). The theoretical values are sizeably larger than the measured ones, and furthermore, this discrepancy is the largest for the models describing tritium binding energy well. On the other hand, as demonstrated in Deltuva and Fonseca [86], the calculated cross-sections for 18 MeV neutrons lie in the middle between the data sets of Seagrave et al. [88] and Debertin et al. [89]. One might thus expect a lack of reliability for the data from Seagrave et al. [88]. As this disagreement is only manifested in the vicinity of the cross-section minima, one is tempted to attribute the discrepancy to a simple underestimation of the experimental error-bars. New precise measurements are required to resolve this issue.

The description of the scattering in the continuum of the $^4$He nucleus, involving three experimentally accessible processes p-$^3$H/n-$^3$He/$^2$H+$^2$H, is the most complicated four-nucleon problem. Nevertheless, an accurate description of this system has been achieved by three different groups, successfully benchmarking their results [15]. The Vilnius-Lisbon group has studied this system extensively in a broad energy region, as well as employing different interaction models [29, 30, 90–92]. One may single out two very challenging energy regions in this system. The first is related to the presence of a $J^\pi = 0^+$ resonant state embedded between the p-$^3$H and n-$^3$He thresholds (see **Figure 1**). Small modifications in the nuclear Hamiltonian affecting the position of this resonant state have huge effects on the calculated cross-sections between the two thresholds. As demonstrated in Lazauskas [93], the majority of the nuclear Hamiltonians fail in this enterprise. Another challenging case is the description of the $^4$He continuum just above the n+$^3$He threshold. In this window, not only the continuum but also elastic n+$^3$He as well as transfer $n + ^3He \rightarrow p + ^3H$ cross-sections are purely reproduced [29, 92]. This feature is determined by the difficulty of describing two relatively narrow ($J^\pi = 0^-$ and $J^\pi = 2^-$) resonant states (see **Figure 1**). One should still explore whether the 3BF models may provide any improvement in describing this region. When increasing energy and moving above the three- and four-nucleon breakup thresholds, in close similarity with p+$^3$He and n+$^3$H systems, description of the scattering cross-sections but also the analyzing powers tends to improve [30, 92].

## 3.4. Five-Nucleon Systems

As mentioned above, the description of a five-nucleon system based on the solution of the Faddeev-Yakubovsky equations represents a considerable technical challenge. Nevertheless, during the last few years, we have achieved a converged solution of the elastic neutron scattering on ⁴He as well as being able to determine the ⁵He resonance position in the complex energy plane. In both cases, the results were based on realistic NN and three-nucleon interaction models.

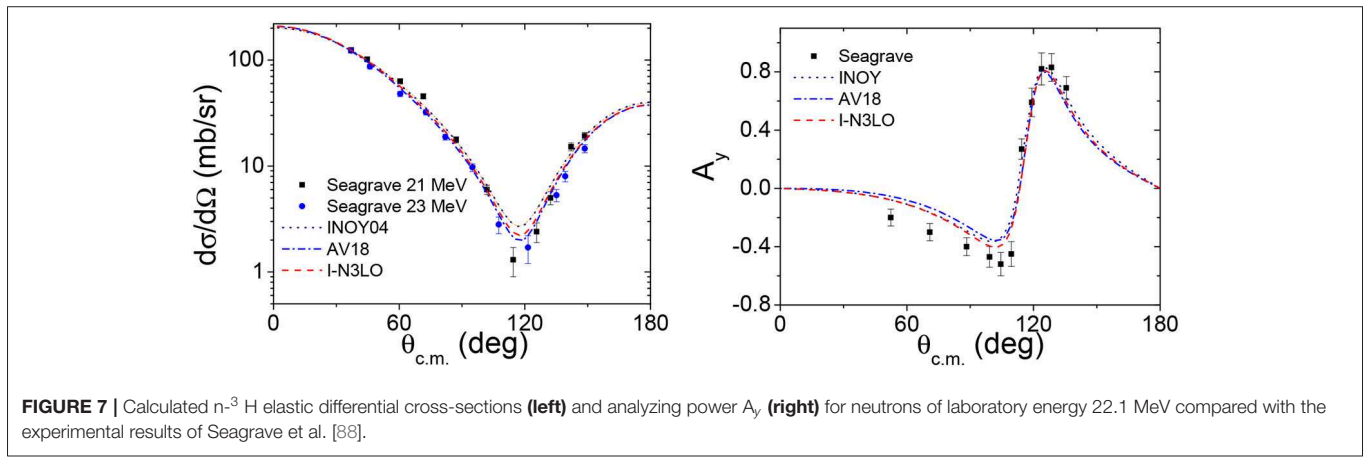

**FIGURE 7 |** Calculated n-³H elastic differential cross-sections **(left)** and analyzing power $A_y$ **(right)** for neutrons of laboratory energy 22.1 MeV compared with the experimental results of Seagrave et al. [88].

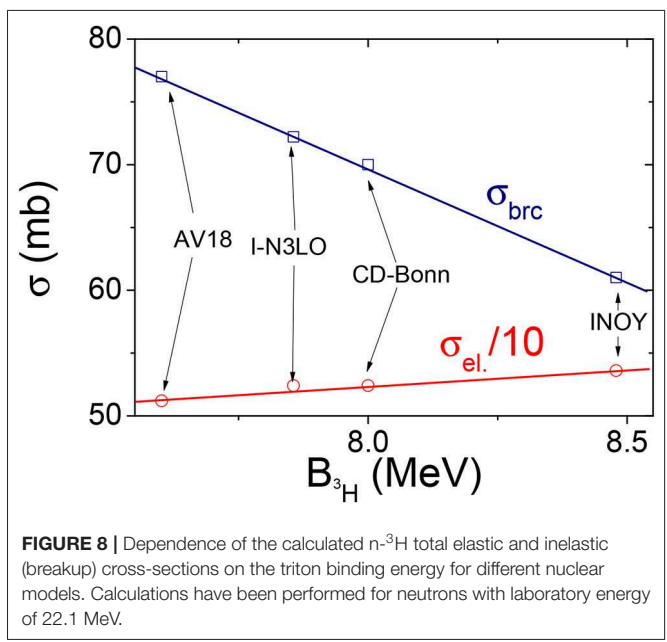

**FIGURE 8 |** Dependence of the calculated n-³H total elastic and inelastic (breakup) cross-sections on the triton binding energy for different nuclear models. Calculations have been performed for neutrons with laboratory energy of 22.1 MeV.

Our results on n-⁴He were in good agreement with some previous calculations based on NCMC techniques [37]. Ideally, one should compare the calculated observables directly with the experimentally available data. However, due to the limited accuracy of the calculations (of order 5% for the phase shifts) and the fact that very accurate phase shift analysis has been carried out on the experimental data for this system, it is practical to analyze the obtained results by comparing the phase shifts.

We present in **Figure 9** our calculated S- and P-wave phaseshifts in the energy region up to 8 MeV. One may see quite a nice description of the scattering observables in the relative S-wave, which also demonstrates a remarkable model independence. In close analogy to the n-³H scattering case, this partial wave is dominated by Pauli repulsion between the neutron projectile and the ones present within the ⁴He target. We would like to note, however, that some model dependence is observed even in S-waves if one compares the phase shifts at very low energy and, in particular, the calculated scattering length. Significant differences are observed between the different theoretical predictions [95] but also between the experimentally measured [96–98] as well as adopted [25, 99, 100] scattering length values. In particular, our calculated values are in conflict with those obtained using GFMC techniques [32], where a scattering length $a(^2S_{1/2}) = 2.4$ fm was found, the same value for AV18 or AV18 supplemented with UIX (or IL2) 3N forces, while our calculations with AV18 give $a(^2S_{1/2}) = 2.96(5)$ fm, whereas for AV18+UIX, we get $a(^2S_{1/2}) = 2.71(7)$. We believe that this difference may be attributable to the lack of accuracy in Nollett et al. [32], as their calculations are not able to reveal any difference in calculated scattering length for AV18 and AV18+UIX Hamiltonians. In contrast, our calculations indicate the presence of a strong correlation between the calculated scattering length and ⁴He binding energy, displayed in **Figure 10**. Therefore, it should be expected that AV18 and AV18+UIX models sizeably differing in predicted ⁴He binding energies should also provide different n-⁴He scattering lengths.

Even more problematic is description of the resonant n-⁴He P-waves. Realistic NN interaction models fail to provide sufficient splitting between the quartet and the doublet P-waves. For the INOY04 model, the situation is even worse: this model significantly lacks attraction in both P-waves. The addition of the UIX 3BF to the AV18 model does not improve description of the n-⁴He P-wave phaseshifts, as was the case for the n-³H and p-³He systems. In contrast, the I-N3LO model, when used in conjunction with the 3BF ($\Lambda = 500$ MeV) from Marcucci et al. [71], turns out to be quite successful in describing both P-waves. The phase shifts of the strongly resonant $^{3/2}P$ channel are reproduced quite well, with only a slight lack of attraction, whereas the $^{1/2}P$ phase shifts are described ideally. The comparison of the results from different interaction models suggests the presence of strong similarities in the n-³H and n-⁴He scattering. There is an apparent correlation between the positions of P-wave resonant states in ⁴H and ⁵He nuclei.

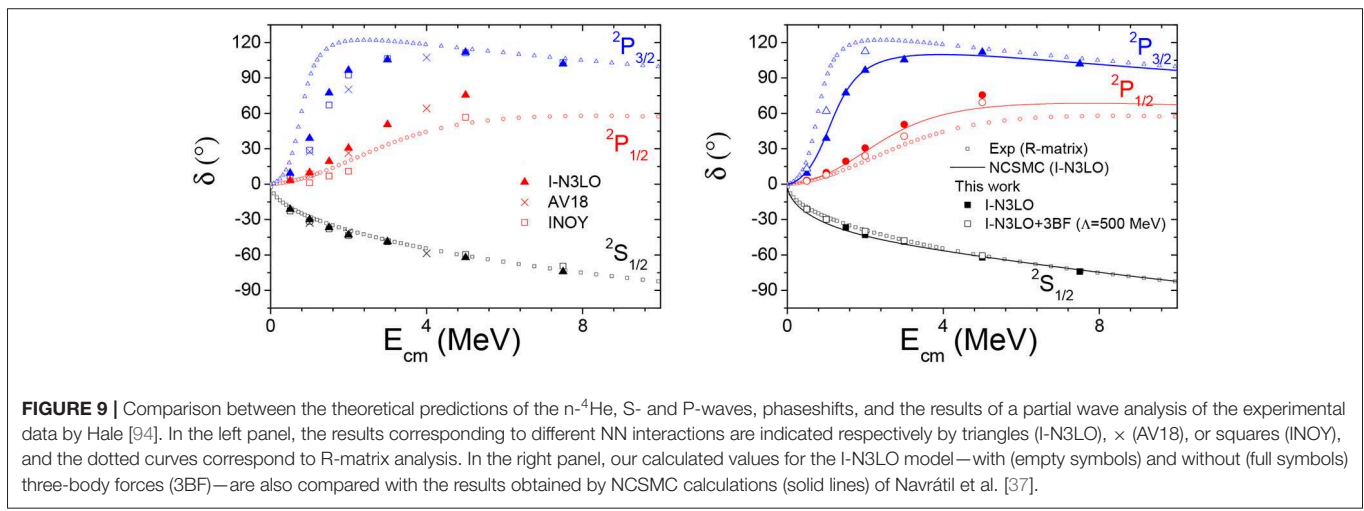

**FIGURE 9 |** Comparison between the theoretical predictions of the n-⁴He, S- and P-waves, phaseshifts, and the results of a partial wave analysis of the experimental data by Hale [94]. In the left panel, the results corresponding to different NN interactions are indicated respectively by triangles (I-N3LO), × (AV18), or squares (INOY), and the dotted curves correspond to R-matrix analysis. In the right panel, our calculated values for the I-N3LO model—with (empty symbols) and without (full symbols) three-body forces (3BF)—are also compared with the results obtained by NCSMC calculations (solid lines) of Navrátil et al. [37].

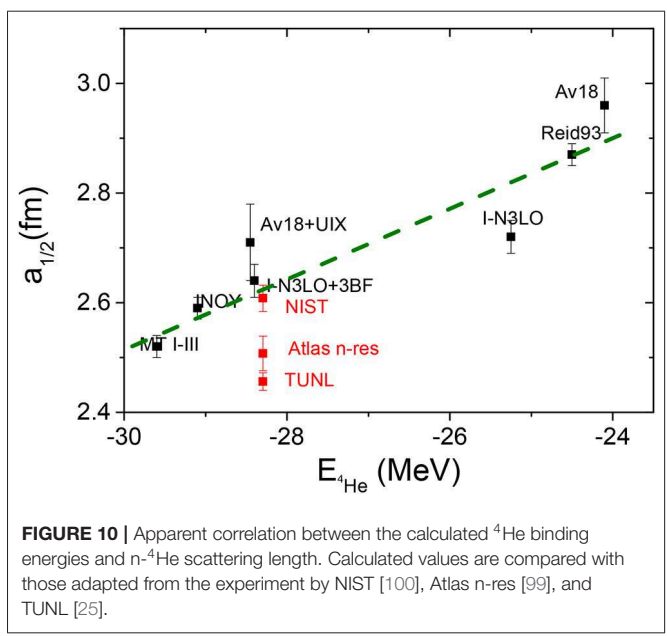

**FIGURE 10 |** Apparent correlation between the calculated ⁴He binding energies and n-⁴He scattering length. Calculated values are compared with those adapted from the experiment by NIST [100], Atlas n-res [99], and TUNL [25].

The ⁵H resonance parameters were first computed ab-initio in Lazaukas et al. [6] with phenomenological and realistic NN interactions. We used two independent methods to locate the resonance positions in the complex energy plane: a variant of the smooth exterior complex scaling method, and the analytic continuation on the coupling constant. The results show remarkable stability with respect to the different tested interactions and support recent experimental findings [101, 102]. The resonance parameters of the $J^\pi = 2^-, 0^-, 1^-$ states in ⁴H, which dominate the low-energy n-³H elastic cross-section, have also been computed and found to be slightly wider than those for ⁵H

($\Gamma_{^4H} \approx 4$ MeV for $\Gamma_{^5H} \approx 2.5$ MeV), advocating for the presence of additional attraction of the 4n with respect to the 3n system. In view of this, any attempt to reproduce the experimental finding of a ⁷H narrow state would be of the highest interest.

## 4. CONCLUSIONS

We have presented some recent results related to the solutions of the Faddeev-Yakubovsky equations in configuration space for nuclei with four or five nucleons obtained with several modern realistic NN and NNN interactions.

Two independent methods to include the Coulomb interaction in the A = 4 scattering system, namely in the p-³He low-energy elastic cross-section, have been compared.

We have discussed in detail the n+³H elastic cross-section in the resonance peak, which constitutes a stumbling block for all realistic NN and 3BF models, even those that most successfully describe the binding energy of A = 3, 4 nuclei.

The mirror reaction p+³He was also presented by computing several observables such as differential cross-sections and analyzing power at $E_p \approx 5$ MeV.

The first results for the five-nucleon system have been considered. They concern the n-⁴He elastic scattering at low energy and the resonance position of ⁵H in the complex energy plane. The n-⁴H scattering displays severe discrepancies in terms of scattering length, both between models and with experimental data. The resonance parameters of ⁵H show great stability with respect to the NN interactions used and are compatible with some of the experimental analyses.

The general conclusion concerning the nuclear interactions is that the I-N3LO NN model used in conjunction with 3BF with a cut-off parameter Λ = 500 MeV provides the most accurate description of the data.

## DATA AVAILABILITY STATEMENT

The raw data supporting the conclusions of this article will be made available by the authors, without undue reservation, to any qualified researcher.

## AUTHOR CONTRIBUTIONS

All authors listed have made a substantial, direct and intellectual contribution to the work, and approved it for publication.

## FUNDING

This work was supported by IN2P3 for the master project Neutron-rich light unstable nuclei.

## ACKNOWLEDGMENTS

We were granted access to the HPC resources of TGCC/IDRIS under the allocation 2018-A0030506006 made by GENCI (Grand Equipement National de Calcul Intensif).

**Conflict of Interest:** The authors declare that the research was conducted in the absence of any commercial or financial relationships that could be construed as a potential conflict of interest.